\documentclass[11pt]{iopart}

\newcommand{\bea}{\begin{eqnarray}}
\newcommand{\eea}{\end{eqnarray}}

\begin{document}

\title{Cosmological post-Newtonian equations from nonlinear perturbation theory}
\author{Hyerim Noh}
\address{Korea Astronomy and Space Science Institute,
         Daejeon 305-348, Republic of Korea}
\ead{hr@kasi.re.kr}
\author{Jai-chan Hwang}
\address{Department of Astronomy and Atmospheric Sciences,
         Kyungpook National University, Daegu 702-701, Republic of Korea}
\ead{jchan@knu.ac.kr}


\begin{abstract}

We derive the basic equations of the cosmological first-order
post-Newtonian approximation from the recently formulated
fully nonlinear and exact cosmological perturbation theory in Einstein's
gravity. Apparently the latter, being exact, should include the
former, and here we use this fact as a new derivation of the former.
The complete sets of equations in both approaches are presented without
fixing the temporal gauge conditions so that we can use the gauge
choice as an advantage. Comparisons between the two approaches are made. Both are potentially important in handling relativistic aspects of nonlinear
processes occurring in cosmological structure formation. We consider an ideal fluid and include the cosmological
constant.

\end{abstract}



\tableofcontents

%
%
\section{Introduction}

The origin and evolution of the large-scale cosmic structures are
important in current scientific cosmology providing links between
theories and observations. Many of the important constraints on the
cosmological models come from matching observations of the
large-scale galaxy distribution and motion, and the cosmic microwave background
radiation with the theories based on the cosmological perturbation
theory.

There are plenty of cosmological situations where the relativistic
nature of Einstein's gravity becomes potentially important. Besides
its geometric nature involving dynamic space and time, the heavy
nonlinear nature of Einstein's gravity also has limited wide
applications of the theory in the later nonlinear evolution stage of
structures in cosmology. Nonlinear aspects of clustering process of
the large-scale structure are mainly studied in the Newtonian
context \cite{Peebles-1980,Zeldovich-Novikov-1983,Bertschinger-1998,
Bernardeau-etal-2002,Kim-etal-2011} except that the background
evolution is provided by Friedmann equations based on Einstein's
gravity often with the cosmological constant. Up to the
leading-order nonlinear perturbation in Einstein's gravity the
success of Newtonian theory is quite remarkable though
\cite{Vishniac-1983,Hwang-Noh-2006,Jeong-etal-2011}.

Recently we notice advents of two new methods in handling the
relativistic and nonlinear aspects of Einstein's gravity in the
Friedmann background cosmology. One is the post-Newtonian (PN)
approximation \cite{Hwang-etal-2008} with previous studies
\cite{Futamase-1988,Tomita-1988,Futamase-1989,Tomita-1991,
Futamase-1993a,Futamase-1993b,Shibata-Asada-1995,
Asada-Futamase-1997,Takada-Futamase-1999}. The other one is the
fully nonlinear and exact perturbation theory in Einstein's gravity
\cite{Hwang-Noh-2013a}.

In comparison to the perturbation approach (Section \ref{sec:PT}) which is weakly nonlinear but fully relativistic, the 1PN approximation (Section \ref{sec:PN}) can be regarded as fully nonlinear but weakly relativistic. Thus, the two approaches are complementary in understanding the relativistic nonlinear evolution stage of the cosmological structures.

In cosmological situations where the nonlinearity as well as
relativistic effects are important we may need full-blown numerical
relativity implemented in cosmology. Such a general relativistic
numerical simulation in cosmology is not currently available yet.
The nonlinear perturbation analysis, being based on perturbative
approach, is not sufficient to handle the genuine nonlinear aspects
of structure formation accompanied with self-organization and
spontaneous formation of structures. In order to handle the
relativistic nonlinear process in cosmology we anticipate the PN
approach is currently practically suitable to implement in numerical
simulation.

Our aim in this work is to {\it derive} the cosmological 1PN equations
from the nonlinear perturbation theory and to show the relation
between the two complementary approaches. The overlap between the 1PN approximation and the {\it linear} perturbation theory was studied in
\cite{Noh-Hwang-2012}. Equations of 1PN approximation, however,
contain perturbations up to the fourth order. As the equations of
fully nonlinear perturbation theory are exact without even
decomposition into background and perturbation, apparently the
perturbation theory should include the 1PN approximation. In this
work we will derive the 1PN equations from the nonlinear
perturbation theory equations. In our way of showing the relation we
will present the complete sets of equations, the gauge strategies,
and solution logics of the two approaches.

\section{Exact and fully nonlinear perturbation theory}
                                             \label{sec:PT}

The perturbation theory assumes all perturbation variables are small
and makes perturbation expansion in those variables. The
relativistic perturbation theory considers fully relativistic
situation, thus applicable in all scales including the superhorizon
scale, and in the early universe where radiation and other
relativistic components have important roles. In Einstein's gravity
the linear perturbation theory has important roles in explaining the
origin and evolution of large-scale structures
\cite{Lifshitz-1946,Harrison-1967,Sachs-Wolfe-1967,Field-Shepley-1968,Nariai-1969,
Bardeen-1980,Bardeen-1988,Mukhanov-1988}. Recently we have
formulated a fully nonlinear and exact perturbation theory
\cite{Hwang-Noh-2013a}. In \cite{Hwang-Noh-2013a} the exact
perturbation equations are also presented without fixing the
temporal gauge condition, thus again in a sort of a gauge-ready
form, see below.

Our metric convention in the perturbation theory is
\cite{Bardeen-1988,Hwang-Noh-2013a} \bea
   & & ds^2 = - \left( 1 + 2 \alpha \right) c^2 d t^2
       - 2 \chi_{i} c d t d x^i
       + a^2 \left( 1 + 2 \varphi \right) \gamma_{ij} d x^i d x^j,
   \label{metric-PT}
\eea where spatial index of $\chi_i$ is raised and lowered by
$\gamma_{ij}$ as the metric; $\gamma_{ij}$ is the comoving part of
spatial metric of the Robertson-Walker spacetime; as we consider a
spatially flat background in the absence of perturbation, we have $\gamma_{ij}
= \delta_{ij}$. Here we {\it assume} $a$ to be a function of $t$
only, and $\alpha$, $\varphi$ and $\chi_i$ are {\it arbitrary} functions
of spacetime. The spatial part of the metric is simple because we
already have taken the spatial gauge condition without losing any
generality to the fully nonlinear perturbation orders
\cite{Bardeen-1988,Hwang-Noh-2013a} and have {\it ignored} the
transverse-tracefree tensor-type perturbation. Our spatial gauge
(congruence) choice is unique in the sense that under our spatial
gauge condition the remaining variables are free from the spatial
gauge mode even to the nonlinear order perturbations
\cite{Hwang-Noh-2013a}; any alternative choice leaves the remnant
spatial gauge mode even after imposing the spatial gauge condition
which should be carefully handled even in the linear order
perturbation. In our spatial gauge together with one of the
fundamental temporal gauge (slicing) conditions to be suggested
below, all the remaining perturbation variables become free from the
gauge mode and can be regarded as gauge-invariant ones even to the
nonlinear perturbation order \cite{Hwang-Noh-2013a}.

It is important to neglect the transverse-tracfree part of the
metric to have the fully nonlinear perturbation formulation in
\cite{Hwang-Noh-2013a}. This is the main {\it assumption} restricting the
potential applications of the formulation. We may still call our
formulation exact and fully nonlinear because, except for ignoring
the transverse-tracefree part, in the basic set of equations we have
not imposed any condition on the amplitude of the perturbation
variables, and also we have not separated the background and
perturbation. Ignoring the transverse-tracefree part of the
perturbation is consistent with the 1PN approximation because
gravitational waves are known to show up from the 2.5PN order
\cite{Chandrasekhar-Esposito-1970}. The transvere-tracefree part, however, can always be handled perturbatively to any desired nonlinear perturbation order; in perturbation theory, the scalar- and vector-type perturbations can work as sources to the gravitational waves from the second order, see Section 8 in \cite{Hwang-Noh-2013a}.

The energy-momentum tensor of an ideal
fluid is \bea
   & & \widetilde T_{ab} = \widetilde \mu \widetilde u_a \widetilde u_b
       + \widetilde p \left( \widetilde u_a \widetilde u_b
       + \widetilde g_{ab} \right),
   \label{Tab}
\eea where tildes indicate covariant quantities; $\widetilde \mu$
and $\widetilde p$ are the covariant energy density and pressure,
respectively, and $\widetilde u_a$ is the normalized fluid
four-vector \cite{Ehlers-1993,Ellis-1971,Ellis-1973}. In the
perturbation theory we may introduce \bea
   & & \widetilde \mu = \mu + \delta \mu, \quad
       \widetilde p = p + \delta p, \quad
       \widetilde u_i \equiv a {v_i \over c},
   \label{fluid-PT}
\eea where the index of $v_i$ is raised and lowered by
$\gamma_{ij}$ as the metric. In the following we will keep
$\widetilde \mu$ and $\widetilde p$ without decomposition.

In \cite{Hwang-Noh-2013a} we also have introduced several different
definitions of the fluid three-velocity. We introduce the coordinate
fluid three-velocity $\overline v^i$ as \bea
   & & {1 \over a} {\overline v^i \over c} \equiv {\widetilde u^i \over \widetilde u^0},
\eea where the index of $\overline v^i$ is raised and lowered by
$\gamma_{ij}$ as the metric; this definition coincides with the one
used in our PN study in \cite{Hwang-etal-2008}, see equation
(\ref{fluid-PN}). Compared with $v_i$ we have \cite{Hwang-Noh-2013a}
\bea
   & & v_i
       = {\widehat \gamma \over {\cal N}} \left[
       \left( 1 + 2 \varphi \right) \overline v_i
       - {c \over a} \chi_i \right]
       \equiv \widehat \gamma \widehat v_i,
   \label{velocities}
\eea where $\widehat \gamma$ is the Lorentz factor \bea
   & & \widehat \gamma
       = \sqrt{ 1 + {v^k v_k \over c^2 (1 + 2 \varphi)} }
       = {1 \over \sqrt{ 1
       - {\widehat v^k \widehat v_k \over c^2 (1 + 2 \varphi)}}}
   \nonumber \\
   & & \qquad
       = {1 \over \sqrt{
       1 - {1 + 2 \varphi \over {\cal N}^2}
       \left( {\overline v^k \over c} - {\chi^k \over a (1 + 2 \varphi)} \right)
       \left( {\overline v_k \over c} - {\chi_k \over a (1 + 2 \varphi)}
       \right)}},
\eea and ${\cal N}$, related to the lapse function ($N \equiv a {\cal N}$), is introduced in
equation (\ref{K-bar-eq}); $\widehat v_i$ is the fluid
three-velocity measured by the Eulerian observer; see the Appendix D
of \cite{Hwang-Noh-2013a}. Although mathematically equivalent, in
this work we will use $\overline v_i$.

We can decompose $\chi_i$, $v_i$ and $\overline v_i$ to scalar- and
vector-type perturbations even to nonlinear perturbation orders as
\bea
   & & \chi_i \equiv c \chi_{,i} + \chi^{(v)}_i, \quad
       v_i \equiv - v_{,i} + v^{(v)}_i, \quad
       \overline v_i \equiv - \overline v_{,i} + \overline v^{(v)}_i,
\eea with the vector-type perturbations satisfying $\chi^{(v)|i}_i
\equiv 0$ and $v^{(v)|i}_i \equiv 0 \equiv \overline v^{(v)|i}_i$; a
vertical bar indicates the covariant derivative based on
$\gamma_{ij}$ as the metric, thus the same as an ordinary derivative
in our present case. Due to the nonlinear relation between $v_i$ and
$\overline v_i$ the scalar- and vector-decompositions for $v_i$ and
$\overline v_i$ do not coincide with each other to the nonlinear
order. To the nonlinear order {\it our} scalar- and vector-type
perturbations are coupled in the equation level.

In order to match with our convention in the PN approach, here we
introduce the dimensions as the following \bea
   \fl [a] = [\gamma_{ij}] = [\widetilde g_{ab}] = [\widetilde u_a] = 1, \quad
       [x^i] = L, \quad
       [\alpha] = [\varphi] = [\chi^i] = [v^i/c] = [\overline v^i/c] = 1, \quad
   \nonumber \\
   \fl
       [\kappa] = T^{-1}, \quad
       [\chi] = T, \quad
       [v/c] = L, \quad
       [\widetilde T_{ab}] = [\widetilde p] = [\widetilde \mu],
       \quad
       [G \widetilde \mu/c^2] = T^{-2},
\eea where $\kappa$ is the perturbed part of the trace of extrinsic
curvature, see equation (\ref{eq1}).

The exact and fully nonlinear perturbation equations, without taking
the temporal gauge (slicing or hypersurface) condition, are the
following \cite{Hwang-Noh-2013a}.

\noindent
Definition of $\kappa$: \bea
   \fl \kappa
       \equiv
       3 {\dot a \over a} \left( 1 - {1 \over {\cal N}} \right)
       - {1 \over {\cal N} (1 + 2 \varphi)}
       \left[ 3 \dot \varphi
       + {c \over a^2} \left( \chi^k_{\;\;,k}
       + {\chi^{k} \varphi_{,k} \over 1 + 2 \varphi} \right)
       \right].
   \label{eq1}
\eea
ADM energy constraint:
\bea
   \fl - {3 \over 2} \left( {\dot a^2 \over a^2}
       - {8 \pi G \over 3 c^2} \widetilde \mu
       - {\Lambda c^2 \over 3} \right)
       + {\dot a \over a} \kappa
       + {c^2 \Delta \varphi \over a^2 (1 + 2 \varphi)^2}
       = {1 \over 6} \kappa^2
       - {4 \pi G \over c^2} \left( \widetilde \mu + \widetilde p \right)
       \left( \widehat \gamma^2 - 1 \right)
   \nonumber \\
   \fl \qquad
       + {3 \over 2} {c^2 \varphi^{,i} \varphi_{,i} \over a^2 (1 + 2 \varphi)^3}
       - {c^2 \over 4} \overline{K}^i_j \overline{K}^j_i.
   \label{eq2}
\eea
ADM momentum constraint:
\bea
   \fl {2 \over 3} \kappa_{,i}
       + {c \over 2 a^2 {\cal N} ( 1 + 2 \varphi )}
       \left( \Delta \chi_i
       + {1 \over 3} \chi^k_{\;\;,ik} \right)
       + {8 \pi G \over c^4} \left( \widetilde \mu + \widetilde p \right)
       {a \widehat \gamma^2 \over {\cal N}}
       \left[ \left( 1 + 2 \varphi \right) \overline v_i
       - {c \over a} \chi_i \right]
   \nonumber \\
   \fl \qquad
       =
       {c \over a^2 {\cal N} ( 1 + 2 \varphi)}
       \Bigg\{
       \left( {{\cal N}_{,j} \over {\cal N}}
       - {\varphi_{,j} \over 1 + 2 \varphi} \right)
       \left[ {1 \over 2} \left( \chi^{j}_{\;\;,i} + \chi_i^{\;,j} \right)
       - {1 \over 3} \delta^j_i \chi^k_{\;\;,k} \right]
   \nonumber \\
   \fl \qquad
       - {\varphi^{,j} \over (1 + 2 \varphi)^2}
       \left( \chi_{i} \varphi_{,j}
       + {1 \over 3} \chi_{j} \varphi_{,i} \right)
       + {{\cal N} \over 1 + 2 \varphi} \nabla_j
       \left[ {1 \over {\cal N}} \left(
       \chi^{j} \varphi_{,i}
       + \chi_{i} \varphi^{,j}
       - {2 \over 3} \delta^j_i \chi^{k} \varphi_{,k} \right) \right]
       \Bigg\}.
   \label{eq3}
\eea
Trace of ADM propagation:
\bea
   \fl - 3 {1 \over {\cal N}}
       \left( {\dot a \over a} \right)^{\displaystyle\cdot}
       - 3 {\dot a^2 \over a^2}
       - {4 \pi G \over c^2} \left( \widetilde \mu + 3 \widetilde p \right)
       + \Lambda c^2
       + {1 \over {\cal {\cal N}}} \dot \kappa
       + 2 {\dot a \over a} \kappa
       + {c^2 \Delta {\cal N} \over a^2 {\cal N} (1 + 2 \varphi)}
   \nonumber \\
   \fl \qquad
       = {1 \over 3} \kappa^2
       + {8 \pi G \over c^2} \left( \widetilde \mu + \widetilde p \right)
       \left( \widehat \gamma^2 - 1 \right)
       - {c \over a^2 {\cal N} (1 + 2 \varphi)} \left(
       \chi^{i} \kappa_{,i}
       + c {\varphi^{,i} {\cal N}_{,i} \over 1 + 2 \varphi} \right)
       + c^2 \overline{K}^i_j \overline{K}^j_i.
   \label{eq4}
\eea
Tracefree ADM propagation:
\bea
   \fl
       \left( {1 \over {\cal N}} {\partial \over \partial t}
       + 3 {\dot a \over a}
       - \kappa
       + {c \chi^{k} \over a^2 {\cal N} (1 + 2 \varphi)} \nabla_k \right)
       \Bigg\{ {c \over a^2 {\cal N} (1 + 2 \varphi)}
   \nonumber \\
   \fl \qquad
       \times \left[
       {1 \over 2} \left( \chi^i_{\;\;,j} + \chi_j^{\;,i} \right)
       - {1 \over 3} \delta^i_j \chi^k_{\;\;,k}
       - {1 \over 1 + 2 \varphi} \left( \chi^{i} \varphi_{,j}
       + \chi_{j} \varphi^{,i}
       - {2 \over 3} \delta^i_j \chi^{k} \varphi_{,k} \right)
       \right] \Bigg\}
   \nonumber \\
   \fl \qquad
       - {c^2 \over a^2 ( 1 + 2 \varphi)}
       \left[ {1 \over 1 + 2 \varphi}
       \left( \nabla^i \nabla_j - {1 \over 3} \delta^i_j \Delta \right) \varphi
       + {1 \over {\cal N}}
       \left( \nabla^i \nabla_j - {1 \over 3} \delta^i_j \Delta \right) {\cal N} \right]
   \nonumber \\
   \fl \qquad
       =
       {8 \pi G \over c^2} \left( \widetilde \mu + \widetilde p \right)
       \Bigg[
       \left( 1 + 2 \varphi \right) {\widehat \gamma^2 \over c^2 {\cal N}^2}
       \left( \overline v^i
       - {c \chi^i \over a ( 1 + 2 \varphi )} \right)
       \left( \overline v_j
       - {c \chi_j \over a ( 1 + 2 \varphi )} \right)
   \nonumber \\
   \fl \qquad
       - {1 \over 3} \delta^i_j \left( \widehat \gamma^2 - 1 \right)
       \Bigg]
       + {c^2 \over a^4 {\cal N}^2 (1 + 2 \varphi)^2} \Bigg[
       {1 \over 2} \left( \chi^{i,k} \chi_{j,k}
       - \chi_{k,j} \chi^{k,i} \right)
   \nonumber \\
   \fl \qquad
       + {1 \over 1 + 2 \varphi} \left(
       \chi^{k,i} \chi_k \varphi_{,j}
       - \chi^{i,k} \chi_j \varphi_{,k}
       + \chi_{k,j} \chi^k \varphi^{,i}
       - \chi_{j,k} \chi^i \varphi^{,k} \right)
   \nonumber \\
   \fl \qquad
       + {2 \over (1 + 2 \varphi)^2} \left(
       \chi^{i} \chi_{j} \varphi^{,k} \varphi_{,k}
       - \chi^{k} \chi_{k} \varphi^{,i} \varphi_{,j} \right) \Bigg]
   \nonumber \\
   \fl \qquad
       - {c^2 \over a^2 (1 + 2 \varphi)^2}
       \left[ {3 \over 1 + 2 \varphi}
       \left( \varphi^{,i} \varphi_{,j}
       - {1 \over 3} \delta^i_j \varphi^{,k} \varphi_{,k} \right)
       + {1 \over {\cal N}} \left(
       \varphi^{,i} {\cal N}_{,j}
       + \varphi_{,j} {\cal N}^{,i}
       - {2 \over 3} \delta^i_j \varphi^{,k} {\cal N}_{,k} \right) \right].
   \nonumber \\
   \label{eq5}
\eea
Covariant energy conservation:
\bea
   \fl \left( {\partial \over \partial t}
       + {1 \over a} \overline v^k \nabla_k \right)
       \widetilde \mu
       + \left( \widetilde \mu + \widetilde p \right)
       {\cal N} \left( 3 H - \kappa \right)
       = - \left( \widetilde \mu + \widetilde p \right)
       \Bigg[ {1 \over a} \left( \overline v^k
       - {c \chi^k \over a ( 1 + 2 \varphi )} \right)_{,k}
   \nonumber \\
   \fl \qquad
       + {3 \varphi_{,k} \over a ( 1 + 2 \varphi )}
       \left( \overline v^k
       - {c \chi^k \over a ( 1 + 2 \varphi )} \right)
       + {1 \over \widehat \gamma}
       \left( {\partial \over \partial t}
       + {1 \over a} \overline v^k \nabla_k \right) \widehat \gamma \Bigg].
   \label{eq6}
\eea
Covariant momentum conservation:
\bea
   \fl {1 \over a \widehat \gamma}
       \left( {\partial \over \partial t}
       + {1 \over a} \overline v^k \nabla_k \right)
       \left\{ {a \widehat \gamma \over {\cal N}}
       \left[ \left( 1 + 2 \varphi \right) \overline v_i
       - {c \over a} \chi_i \right] \right\}
   \nonumber \\
   \fl \qquad
       + {1 \over \widetilde \mu + \widetilde p}
       \left\{
       {c^2 {\cal N} \over a \widehat \gamma^2} \widetilde p_{,i}
       + {1 \over {\cal N}}
       \left[ \left( 1 + 2 \varphi \right) \overline v_i
       - {c \over a} \chi_i \right]
       \left( {\partial \over \partial t}
       + {1 \over a} \overline v^k \nabla_k \right) \widetilde p
       \right\}
   \nonumber \\
   \fl \qquad
       = - {c^2 \over a} {\cal N}_{,i}
       - {c \over a^2 {\cal N}}
       \left[ \left( 1 + 2 \varphi \right) \overline v^k
       - {c \over a} \chi^k \right]
       \left( {\chi_k \over 1 + 2 \varphi} \right)_{,i}
       + \left( 1 - {1 \over \widehat \gamma^2} \right)
       {c^2 {\cal N} \varphi_{,i} \over a ( 1 + 2 \varphi )}.
   \label{eq7}
\eea These were derived in Section 3 of \cite{Hwang-Noh-2013a}; here
we use $\overline v_i$ as the fluid three-velocity instead of $v_i$
used in \cite{Hwang-Noh-2013a} or $\widehat v_i$ used in \cite{Hwang-Noh-2013b}. With ${\cal N}$ and
$\overline{K}^i_j \overline{K}^j_i$ given as \bea
   \fl {\cal N} \equiv \sqrt{ 1 + 2 \alpha
       + {\chi^k \chi_k \over a^2 ( 1 + 2 \varphi )}}, \quad
       \overline{K}^i_j \overline{K}^j_i
       = {1 \over a^4 {\cal N}^2 (1 + 2 \varphi)^2}
       \Bigg\{
       {1 \over 2} \chi^{i,j} \left( \chi_{i,j} + \chi_{j,i} \right)
       - {1 \over 3} \chi^i_{\;\;,i} \chi^j_{\;\;,j}
   \nonumber \\
   \fl
       - {4 \over 1 + 2 \varphi} \left[
       {1 \over 2} \chi^i \varphi^{,j} \left(
       \chi_{i,j} + \chi_{j,i} \right)
       - {1 \over 3} \chi^i_{\;\;,i} \chi^j \varphi_{,j} \right]
       + {2 \over (1 + 2 \varphi)^2} \left(
       \chi^{i} \chi_{i} \varphi^{,j} \varphi_{,j}
       + {1 \over 3} \chi^i \chi^j \varphi_{,i} \varphi_{,j} \right) \Bigg\},
   \nonumber \\
   \label{K-bar-eq}
\eea equations (\ref{eq1})-(\ref{eq7}) are the complete set of exact
and fully nonlinear perturbation equations valid for the scalar- and
vector-type perturbations assuming an ideal fluid in a flat
background; $\Lambda$ is the cosmological constant. Notice that we
have not separated the background order equations. We {\it only}
have assumed that $a$ is a function of $t$. In this sense the above
set of equations is {\it exact}.

In the above set of equations we have not taken the temporal gauge
condition yet. As the temporal gauge condition we can impose any one
of the following conditions \cite{Hwang-Noh-2013a} \bea
   & & {\rm comoving \; gauge:}              \hskip 2.33cm     \widehat v \equiv 0 \;\; {\rm or} \;\; v \equiv 0,
   \nonumber \\
   & & {\rm zero\!-\!shear \; gauge:}        \hskip 1.94cm     \chi \equiv 0,
   \nonumber \\
   & & {\rm uniform\!-\!curvature \; gauge:} \hskip .50cm      \varphi \equiv 0,
   \nonumber \\
   & & {\rm uniform\!-\!expansion \; gauge:} \hskip .47cm      \kappa \equiv 0,
   \nonumber \\
   & & {\rm uniform\!-\!density \; gauge:}   \hskip 1.00cm      \delta \equiv 0,
   \label{temporal-gauges-NL}
\eea or combinations of these to all perturbation orders; we can
also impose different gauge conditions to different perturbation
orders. With the imposition of any one of these slicing conditions the
remaining perturbation variables are free from the remnant (spatial
and temporal) gauge mode, and have unique gauge-invariant
combinations even to the nonlinear order
\cite{Bardeen-1988,Noh-Hwang-2004,Hwang-Noh-2013a}.

As long as we take perturbation approach the gauge issues (gauge transformation properties and gauge-invariant combinations) can be handled order by order to any nonlinear order (i.e., to the fully nonlinear order). For example, by decomposing the perturbation variable as $\chi \equiv \chi^{(1)} + \chi^{(2)} + \chi^{(3)} + \dots$ with the numerical indices inside the parenthesis indicating the perturbation orders, the zero-shear gauge can be imposed as $\chi^{(i)} = 0$ to each perturbation order to the fully nonlinear order or equivalently $\chi= 0$ as an exact zero-shear condition. The gauge issues to the nonlinear order perturbations in concrete forms are presented in Section VI.C of \cite{Noh-Hwang-2004} and Section 2 of \cite{Hwang-Noh-2013a}.

It is important to notice that the comoving gauge condition is imposed on $v_i$ or $\widehat v_i$. Even to the linear order, under our congruence
(spatial gauge) condition, $\overline v$ is slicing (temporal gauge)
condition independent; to the linear order we have $\overline v = v
- \chi/a \equiv v_\chi$ which is already gauge-invariant (under our
congruence condition). To the nonlinear order we can regard $\widehat v \equiv 0$ or $v \equiv 0$ as the comoving gauge condition. For pure scalar-type perturbation we have $\widehat v = 0$ implies $v = 0$ and {\it vice versa}; in the presence of vector-type perturbation, $\widehat v = 0$ differs from $v = 0$ from the third order in perturbation \cite{Hwang-Noh-2013a}.

%
%
\section{1PN approximation}
                                          \label{sec:PN}

The PN approach abandons the geometric spirit of Einstein's gravity
and provides the relativistic effects as correction terms in the
well known Newtonian equations. That is, the PN approach recovers
the concept of absolute space and absolute time. In this way it
provides the relativistic effects in the forms of correction terms
to the well known Newtonian equations, thus enabling us to use
simpler conventional (numerical) treatment. The corrections are made
based on an expansion in the dimensionless quantity ${GM \over
Rc^2}$ which is of the same order as ${v^2 \over c^2}$ in motions
supported by gravity; $M$, $R$ and $v$ are characteristic mass,
dimension and velocity, respectively, of the system we are
considering. The PN order $n$ is the same as expansion up to $({GM
\over Rc^2})^n \sim ({v \over c})^{2n}$. The PN equations are
applicable to weakly relativistic situation with ${GM \over R c^2}
\ll 1$, and in the subhorizon scale, but are fully nonlinear.

In this work we will consider the first-order PN (1PN) approximation
with $n = 1$.
Chandrasekhar \cite{Chandrasekhar-1965} has derived the 1PN hydrodynamic equations in the Minkowski background in certain gauge condition, see equation (\ref{temporal-gauges-PN}). In \cite{Hwang-etal-2008} we have derived 1PN hydrodynamic equations in the cosmological background in a gauge-ready form where the temporal gauge condition is left (unfixed) as an option for later use, see equation (\ref{Gauge-PN}). In \cite{Hwang-etal-2008} we considered a fluid with general pressure, anisotropic stress and flux; we also have considered the presence of cosmological constant and have shown that the proper 1PN approximation demands the flat background. In this work we ignore the anisotropic stress and the flux terms. Our aim in this section is to re-derive the 1PN hydrodynamic equations from the nonlinear perturbation equations summarized in the previous section. Here we closely follow the conventions used in our 1PN formulation \cite{Hwang-etal-2008}.

To the 1PN order Chandrasekhar's metric convention adopted to the cosmological background is \cite{Chandrasekhar-1965,Chandrasekhar-Nutku-1969,Hwang-etal-2008}
\bea
   \fl
       ds^2 = - \left[ 1 - {1 \over c^2} 2 U
       + {1 \over c^4} \left( 2 U^2 - 4 \Phi \right) \right] c^2 d t^2
       - {1 \over c^3} 2 a P_i c dt d x^i
       + a^2 \left( 1 + {1 \over c^2} 2 V \right) \gamma_{ij} d x^i d x^j,
   \nonumber \\
   \label{metric-PN}
\eea where the spatial index of $P_i$ is raised and lowered by
$\gamma_{ij}$, the comoving spatial part of the Robertson-Walker metric; as we consider a spatially {\it flat} background, we may set $\gamma_{ij} = \delta_{ij}$.
Here, $a(t)$ can be regarded as the cosmic
scale factor. Similarly as in the metric of perturbation theory in
equation (\ref{metric-PT}), in the $\widetilde g_{ij}$ part we
already have taken spatial gauge conditions without losing any
generality, and have ignored the tensor-type perturbation: see
Section 6 in \cite{Hwang-etal-2008}. Comparing the two metrics in
equations (\ref{metric-PT}) and (\ref{metric-PN}), to the 1PN order
we can identify \bea
   & & \alpha = - {1 \over c^2} \left[ U
       - {1 \over c^2} \left( U^2 - 2 \Phi \right) \right], \quad
       \varphi = {1 \over c^2} V, \quad
       \chi_i = {1 \over c^3} a P_i.
   \label{PT-PN-metric}
\eea

The 1PN energy-momentum tensor is presented in equation (21) of
\cite{Hwang-etal-2008}. With vanishing flux and anisotropic stress,
the energy-momentum tensor is given in equation (\ref{Tab}). Comparing the energy-momentum tensors and the four-vectors in the two formulations, to the
1PN order we can identify \bea
   & &
       \widetilde \mu \equiv \widetilde \varrho c^2
       \left( 1 + {1 \over c^2} \widetilde \Pi \right), \quad
       \overline v^i \equiv {\bf v}, \quad
       \widetilde u^i
       \equiv {1 \over c} {1 \over a} \overline{v}^i \widetilde u^0,
   \label{fluid-PN}
\eea where $\widetilde \Pi$ is associated with the internal energy
\cite{Chandrasekhar-1965}; $\overline{v}_i$ is the same as $v_i$
used in \cite{Hwang-etal-2008}. From equation (\ref{velocities}) we
have \bea
   & &
       v_i = \overline v_i
       + {1 \over c^2} \left[
       \overline{v}_i
       \left( {1 \over 2} \overline{v}^2 + U + 2 V \right)
       - P_i \right]
       = \left( 1 + {1 \over c^2} {1 \over 2} \widehat v^2 \right) \widehat v_i,
\eea where $\overline v^2 \equiv \overline v^k \overline v_k$ and
$\widehat v^2 \equiv \widehat v^k \widehat v_k$. The dimensions are
the following \bea
   & & [U] = [V] = [\widetilde \Pi] = c^2, \quad
       [P_i] = c^3, \quad
       [\Phi] = c^4, \quad
       [v^i] = [\overline v^i] = c,
   \nonumber \\
   & &
       [\widetilde p] = [\widetilde \varrho c^2], \quad
       [G \widetilde \varrho] = T^{-2}.
\eea

Now, using the identifications between the two approaches made in
equations (\ref{PT-PN-metric}) and (\ref{fluid-PN}), we can {\it derive}
the 1PN equations from the nonlinear perturbation equations in
(\ref{eq1})-(\ref{eq7}). Equation (\ref{eq1}) gives \bea
   & &
       \kappa = - {1 \over c^2} \left( 3 {\dot a \over a} U
       + 3 \dot V
       + {1 \over a} P^k_{\;\;|k} \right).
   \label{kappa-PN}
\eea
Equation (\ref{eq5}) to 1PN order gives \bea
   & & V = U.
   \label{U-V}
\eea Using these, from equations (\ref{eq6}), (\ref{eq7}), (\ref{eq4}) and (\ref{eq3}), respectively, we can derive \bea
   \fl
       {1 \over a^3} \left( a^3 \widetilde \varrho \right)^{\displaystyle\cdot}
       + {1 \over a} \left( \widetilde \varrho \overline{v}^i \right)_{|i}
       = - {1 \over c^2} \Bigg[ \widetilde \varrho \left( {\partial \over \partial t}
       + {1 \over a} \overline{\bf v} \cdot \nabla \right)
       \left( {1 \over 2} \overline{v}^2 + 3 U + \widetilde \Pi \right)
       + \left( 3 {\dot a \over a}
       + {1 \over a} \nabla \cdot \overline{\bf v} \right) \widetilde p \Bigg],
   \nonumber \\
   \label{E-conserv-PN} \\
   \fl
       {1 \over a} \left( a \overline{v}_i \right)^{\displaystyle\cdot}
       + {1 \over a} \overline{v}_{i|k} \overline{v}^k
       - {1 \over a} U_{,i}
       + {1 \over a} {\widetilde p_{,i} \over \widetilde \varrho}
       = {1 \over c^2} \Bigg[
       {1 \over a} \overline{v}^2 U_{,i}
       + {2 \over a} \left( \Phi - U^2 \right)_{,i}
       + {1 \over a} \left( a P_i \right)^{\displaystyle\cdot}
   \nonumber \\
   \fl \qquad
       + {1 \over a} \overline{v}^k
       \left( P_{i|k} - P_{k|i} \right)
       + {1 \over a} \left( \overline{v}^2 + 4 U + \widetilde \Pi + {\widetilde p \over \widetilde \varrho}
       \right) {\widetilde p_{,i} \over \widetilde \varrho}
       - \overline{v}_i \left( {\partial \over \partial t}
       + {1 \over a} \overline{\bf v} \cdot \nabla \right)
       \left( {1 \over 2} \overline{v}^2 + 3 U \right)
   \nonumber \\
   \fl \qquad
       - \overline{v}_i {1 \over \widetilde \varrho} \left( {\partial \over \partial t}
       + {1 \over a} \overline{\bf v} \cdot \nabla \right) \widetilde p
       \Bigg],
   \label{Mom-conserv-PN} \\
   \fl
       {\Delta \over a^2} U
       + 4 \pi G \left( \widetilde \varrho - \varrho \right)
       = - {1 \over c^2} \Bigg\{
       {1 \over a^2} \left[
       2 \Delta \Phi
       - 2 U \Delta U
       + \left( a P^i_{\;\;|i} \right)^{\displaystyle\cdot}
       \right]
       + 3 \ddot U
       + 9 {\dot a \over a} \dot U
       + 6 {\ddot a \over a} U
   \nonumber \\
   \fl \qquad
       + 8 \pi G \left[ \widetilde \varrho \overline{v}^2
       + {1 \over 2} \left( \widetilde \varrho \widetilde \Pi - \varrho \Pi \right)
       + {3 \over 2} \left( \widetilde p - p \right) \right]
       \Bigg\},
   \label{Raychaudhury-eq} \\
   \fl
       0 = {1 \over a^2} \left( P^k_{\;\;|ki}
       - \Delta P_i \right)
       - 16 \pi G \widetilde \varrho \overline{v}_i
       + {4 \over a} \left( \dot U + {\dot a \over a} U \right)_{,i}.
   \label{Mom-constr-PN}
\eea Terms on the left-hand-side provide the Newtonian (0PN) limit,
and the ones in the right-hand-side are 1PN contributions. Notice
that the 1PN terms include up to fourth-order perturbations.

These are the same as cosmological 1PN equations presented in
\cite{Hwang-etal-2008} for vanishing anisotropic stress and flux; in
the Minkowski background and under a certain gauge condition, see
\cite{Chandrasekhar-1965}. These follow from the
energy-conservation, momentum-conservation, trace of ADM propagation
($\widetilde G^0_0 - \widetilde G^i_i = 2 \widetilde R^0_0$), and
momentum-constraint ($\widetilde G^0_i$) equations, respectively;
for the general case with the anisotropic stress, see equations
(114), (115), (119) and (120) in \cite{Hwang-etal-2008}. In the PN
approximation, the energy constraint equation ($2 \widetilde G^0_0 =
\widetilde R^0_0 - \widetilde R^i_i$) in equation (\ref{eq2}) gives
0PN part of equation (\ref{Raychaudhury-eq}); see equation (8) and
equations (80) and (87) in \cite{Hwang-etal-2008}.

Our cosmological PN approach {\it assumes} a flat cosmological background, but is valid in the presence of the cosmological constant. Equations for
the background are \bea
   & & {\ddot a \over a}
       = - {4 \pi G \over 3} \varrho \left[
       1 + {1 \over c^2} \left( \Pi + 3 {p \over \varrho} \right) \right]
       + {\Lambda c^2 \over 3},
   \nonumber \\
   & &
       {\dot a^2 \over a^2}
       = {8 \pi G \over 3} \varrho \left( 1 + {1 \over c^2} \Pi \right)
       + {\Lambda c^2 \over 3}, \quad
       \dot \varrho
       = - 3 {\dot a \over a} \varrho,
   \label{BG-eqs}
\eea and $\dot \mu + 3 (\dot a /a) ( \mu + p ) = 0$ with $\dot \Pi +
3 (\dot a /a) p/\varrho = 0$; see Section 4.1 in
\cite{Hwang-etal-2008}. These background order equations based on
Einstein's gravity were {\it subtracted} in deriving the PN
equations; see Section 3.2 in \cite{Hwang-etal-2008}. This is
related to the fact that self-consistent treatment of cosmological
world model is {\it not} possible in Newton's gravity. Without a guide by
Einstein's gravity \cite{Friedmann-1922}, the spatially homogeneous
and isotropic cosmological world model based on Newton's gravity is
known to be incomplete and indeterminate
\cite{Layzer-1954,Lemons-1988}.

In order to solve the 1PN equations it will be convenient to have
the equations for $U$, $\dot U$ and $\ddot U$ to the 0PN order
\cite{Shibata-Asada-1995}. For $U$, equation (\ref{Raychaudhury-eq})
gives \bea
   & & {\Delta \over a^2} U
       = - 4 \pi G \left( \widetilde \varrho - \varrho \right),
   \label{U-0PN}
\eea valid to the 0PN order. By taking a divergence of equation (\ref{Mom-constr-PN}), and using equation (\ref{Raychaudhury-eq}), we have \bea
   & & {\Delta \over a^2} \dot U
       = 4 \pi G \left[ {\dot a \over a} \left( \widetilde \varrho - \varrho \right)
       + {1 \over a} \left( \widetilde \varrho \overline{v}^i \right)_{|i} \right],
   \label{dot-U}
\eea valid to the 0PN order. By taking a time derivative, and using equations (\ref{E-conserv-PN})-(\ref{Raychaudhury-eq}), we have \bea
   \fl {\Delta \over a^2} \ddot U
       = 4 \pi G \Bigg\{ \left( {\ddot a \over a}
       - 2 {\dot a^2 \over a^2} \right)
       \left( \widetilde \varrho - \varrho \right)
       - {1 \over a^2} \left[ 4 \dot a \widetilde \varrho \overline{v}^i
       + \left( \widetilde \varrho \overline{v}^i \overline{v}^k \right)_{|k}
       - \widetilde \varrho U^{,i}
       + \widetilde p^{,i}
       \right]_{|i} \Bigg\},
   \label{ddot-U}
\eea valid to the 0PN order.

In the above 1PN equations the spatial gauge conditions were fixed
in the same way as in the perturbation theory, but we have not fixed
the temporal (slicing) condition yet. The gauge transformation property in the 1PN approximation was thoroughly studied in Section 6 of \cite{Hwang-etal-2008}, and in the following we briefly review it. In our 1PN metric convention
in equation (\ref{metric-PN}) we already have taken spatial gauge
condition by setting $g_{ij} = a^2 ( 1 + c^{-2} 2 V ) \delta_{ij}$;
see Section 6 in \cite{Hwang-etal-2008}. Under the remaining gauge
transformation $\widehat x^a = x^a + \xi^a (x^e)$ with \bea
   & & \xi^0 = {1 \over c} \xi^{(2)0} + {1 \over c^3} \xi^{(4)0},
\eea we can set $\xi^{(2)0} = 0$ without losing any generality: see
equation (173) in \cite{Hwang-etal-2008}. In this case variables $U$
and $V$ are gauge-invariant, and we have [see equations (175) and
(176) in \cite{Hwang-etal-2008}] \bea
   & & \widehat P_i = P_i - {1 \over a} \xi^{(4)0}_{\;\;\;\;\;\;,i}, \quad
       \widehat \Phi = \Phi + {1 \over 2} \dot \xi^{(4)0}.
\eea Thus, in the PN approach we have freedom to impose the temporal
gauge (slicing or hypersurface) condition on $P^i_{\;\;|i}$ or
$\Phi$; fixing $\Phi = 0$ in all coordinates leaves the remnant
gauge mode $\xi^{(4)0} ({\bf x})$, whereas setting $P^i_{\;\;|i} =
0$ in all coordinate as the gauge condition completely removes the
gauge mode. A combination $2 \Phi_{,i} + ( a P_i
)^{\displaystyle\cdot}$ is gauge-invariant. The variables
$\widetilde \varrho$ and $\widetilde \Pi$ are not fixed individually
under the gauge transformation, and varies as $\widehat {\widetilde
\varrho} = \widetilde \varrho + \varrho^{(2)}/c^2$ and $\widehat
{\widetilde \Pi} = \widetilde \Pi - \varrho^{(2)}/\varrho$ where
$\varrho^{(2)}$ is an undetermined transformation function, see
equation (192) in \cite{Hwang-etal-2008}. Thus a combination
$\widetilde \varrho (1 + \widetilde \Pi/c^2)$ is gauge invariant;
$\overline{v}_i$ is also gauge invariant: see equation (193) in
\cite{Hwang-etal-2008}. Using the above gauge transformation
properties we can show that equations
(\ref{E-conserv-PN})-(\ref{Mom-constr-PN}) are invariant under the
gauge transformation to the appropriate PN orders.

In \cite{Hwang-etal-2008} we have introduced a {\it general} temporal gauge
condition as \bea
   & & {1 \over a} P^i_{\;\;|i} + n \dot U + m {\dot a \over a} U = 0,
   \label{Gauge-PN}
\eea where $n$ and $m$ can be arbitrary real numbers. Several temporal gauge conditions used in the literature are \bea
   & & {\rm Harmonic \; gauge:}              \hskip 2.28cm
       n= 4, \; m = {\rm arbitrary},
   \nonumber \\
   & & {\rm Chandrasekhar's \; gauge:}       \hskip 1.07cm
       n=3, \; m= {\rm arbitrary},
   \nonumber \\
   & & {\rm Uniform\!-\!expansion \; gauge:} \hskip .4cm
       n = 3 = m,
   \nonumber \\
   & & {\rm Transverse\!-\!shear \; gauge:}  \hskip .82cm
       n = 0 = m.
   \label{temporal-gauges-PN}
\eea Compared with the fundamental temporal gauge conditions of the
perturbation theory in equation (\ref{temporal-gauges-NL}), the
uniform-expansion gauge is the same as $\kappa = 0$ in the
perturbation theory; the Chandrasekhar's gauge is the same as
$\kappa + (m-3) (\dot a/a) \varphi = 0$ in the perturbation theory
which is a fine gauge condition without remnant gauge mode. In the
cosmological PN approach, \cite{Shibata-Asada-1995} also took $n =
3$ as their slicing condition. The transverse-shear gauge is the
same as the zero-shear gauge with $\chi = 0$ (or $\chi^i_{\;\; |i} =
0$). The Harmonic gauge condition can be written as $\dot \alpha + H
\alpha  + \kappa = 0$ which leaves heavy remnant gauge modes in the
perturbation theory even to the linear order (see the Appendix in
\cite{Hwang-1993}), but not in the case of 1PN approximation
\cite{Hwang-etal-2008}. In the PN approximation the comoving gauge
($v^i_{\;\;|i} = 0$), the uniform-curvature gauge ($\varphi = 0$),
and the uniform-density gauge ($\delta \equiv 0$) in the
perturbation theory, corresponding to $\overline{v}^i_{\;\;|i} \equiv 0$,
$V \equiv 0$, and $\widetilde \varrho - \varrho \equiv 0$, respectively, are
{\it not} available.

Equation (\ref{Raychaudhury-eq}) shows that the harmonic gauge
condition makes the Laplacian operator ${\Delta \over a^2}$ in the
0PN limit replaced by a d'Alembertian operator ${\Delta \over a^2} -
{\partial^2 \over c^2 \partial t^2}$ by the 1PN correction terms,
thus making the Poisson's equation in 0PN limit to become a wave
equation with the propagation speed $c$ by the 1PN correction.
Examination of equation (\ref{Raychaudhury-eq}) shows that, under
the general gauge condition in equation (\ref{Gauge-PN}), the
propagation speed of the gravitational potential $U$ is
\bea
   & & {c \over \sqrt{n-3}},
\eea see equation (213) of \cite{Hwang-etal-2008}. Thus, the uniform-expansion gauge and the Chandrasekhar's gauge leave the action-at-a-distance nature of the Poisson's equation, and the transverse-shear gauge makes equation (\ref{Raychaudhury-eq}) to be no longer a wave
equation. Although the propagation speed of the gravitational
potential depends on the gauge choice, the propagation speed of the (physical)
Weyl tensor naturally does not depend on the gauge choice and is
always $c$: see Section 7 of \cite{Hwang-etal-2008}.
Exact analogy can be found in the electromagnetism where the propagation speed of the field potential depends on the gauge choice, whereas the the propagation speed of the (physical) field strength is always the speed of light, see \cite{Jackson-2002}.

Under the gauge condition in equation (\ref{Gauge-PN}), equation
(\ref{Mom-constr-PN}) becomes \bea
   & & {\Delta \over a^2} P_i
       = {1 \over a} \left[ \left( 4 - n \right) \dot U
       + \left( 4 - m \right) {\dot a \over a} U \right]_{,i}
       - 16 \pi G \widetilde \varrho \overline{v}_i.
   \label{Mom-constr-PN-2}
\eea The temporal gauge condition corresponds to fixing $n$ and $m$.
As $U$ is gauge-invariant to 1PN order, independently of the values
of $n$ and $m$, using equation (\ref{Mom-constr-PN-2}) all variables
in our set of equations are free from the gauge mode, and can be
regarded as gauge invariant.

By introducing a gauge-invariant combination \bea
   & & {\cal U}_{,i}
       \equiv U_{,i}
       + {1 \over c^2} \left[ 2 \Phi_{,i}
       + \left( a P_i \right)^{\displaystyle\cdot}
       \right],
   \label{cal_U}
\eea the $\Phi$ and $(a P_i)^{\displaystyle\cdot}$ terms in
the right-hand-sides of equations (\ref{Mom-conserv-PN}) and
(\ref{Raychaudhury-eq}) can be absorbed to the $U$ terms in the
left-hand-sides by replacing $U$ to ${\cal U}$. Equations
(\ref{Mom-conserv-PN}) and (\ref{Raychaudhury-eq}) become \bea
   & &
       {1 \over a} \left( a \overline{v}_i \right)^{\displaystyle\cdot}
       + {1 \over a} \overline{v}_{i|k} \overline{v}^k
       - {1 \over a} {\cal U}_{,i}
       + {1 \over a} {\widetilde p_{,i} \over \widetilde \varrho}
       = {1 \over c^2} \Bigg[
       {1 \over a} \left( \overline{v}^2 - 4 U \right) U_{,i}
       + {2 \over a} \overline{v}^k P_{[i|k]}
   \nonumber \\
   & & \qquad
       + {1 \over a} \left( \overline{v}^2 + 4 U + \widetilde \Pi
       + {\widetilde p \over \widetilde \varrho}
       \right) {\widetilde p_{,i} \over \widetilde \varrho}
       - \overline{v}_i \left( {\partial \over \partial t}
       + {1 \over a} \overline{\bf v} \cdot \nabla \right)
       \left( {1 \over 2} \overline{v}^2 + 3 U \right)
   \nonumber \\
   & & \qquad
       - \overline{v}_i {1 \over \widetilde \varrho}
       \left( {\partial \over \partial t}
       + {1 \over a} \overline{\bf v} \cdot \nabla \right) \widetilde p
       \Bigg],
   \label{Mom-conserv-PN-2} \\
   & &
       {\Delta \over a^2} {\cal U}
       + 4 \pi G \left( \widetilde \varrho - \varrho\right)
       = - {1 \over c^2} \Bigg\{
       3 \ddot U
       + 9 {\dot a \over a} \dot U
       + 6 {\ddot a \over a} U
   \nonumber \\
   & & \qquad
       + 8 \pi G \left[
       \widetilde \varrho \overline{v}^2
       + \left( \widetilde \varrho - \varrho \right) U
       + {1 \over 2} \left( \widetilde \varrho \widetilde \Pi - \varrho \Pi \right)
       + {3 \over 2} \left( \widetilde p - p \right) \right]
       \Bigg\}.
   \label{Raychaudhury-eq-2}
\eea

Analysis can be proceeded as follows. The fluid variables
$\widetilde p$ and $\widetilde \Pi$ should be given by the equation
of state according to the system we are modeling; for example, for
pressureless dust or cold dark matter, we may set $\widetilde p = 0
= \widetilde \Pi$. Choose the gauge condition, thus fixing $n$ and
$m$ in equation (\ref{Mom-constr-PN-2}). We can determine
$\widetilde \varrho$, $\overline{v}_i$ and ${\cal U}$ to 1PN order
by solving equations (\ref{E-conserv-PN}), (\ref{Mom-conserv-PN-2})
and (\ref{Raychaudhury-eq-2}). In order to determine $U$, $\dot U$,
$\ddot U$, and $P_i$ in the right-hand-sides of these equations to
the 0PN order, we can solve equations (\ref{U-0PN}), (\ref{dot-U}),
(\ref{ddot-U}), and (\ref{Mom-constr-PN-2}) respectively; inverting
the Laplacian by volume integration in these equations would be time
consuming in numerical implementation. The background evolution
should be determined by equation (\ref{BG-eqs}) simultaneously.

We may identify cosmological situations where the cosmological PN approach might have important applications. The current cosmological paradigm favors a model where the large-scale structures are in the linear stage, whereas small-scale structures are apparently in fully nonlinear stage. The standard strategy is to assume that the small-scale nonlinear structures can be fully handled by the Newtonian gravity. In the galactic and cluster scales we have the general relativistic measure ${GM \over Rc^2} \sim {v^2 \over c^2} \sim 10^{-6} - 10^{-4}$, thus small but nonvanishing, thus indeed the 1PN (weakly relativistic) assumption is quite sufficiently valid. We believe the 1PN approach would be relevant to estimate the general relativistic effects in the nonlinear clustering processes of the galaxy cluster-scale and the large-scale structures.

%
%
\section{0PN approximation}
                                          \label{sec:0PN}

We can view the Newtonian limit as the 0PN approximation; see Sec.\
4.2 of \cite{Hwang-etal-2008}. Indeed, the 1PN approximation studied
in the previous section properly contains the Newtonian hydrodynamic
equations at the 0PN level; in the Minkowski background, see
\cite{Chandrasekhar-1965}. In order to have the Newtonian
hydrodynamic equations as the 0PN limit, we should take the 0PN
metric as \bea
   & & ds^2 = - \left( 1 - {1 \over c^2} 2 U \right) c^2 d t^2
       + a^2 \left( 1 + {1 \over c^2} 2 V \right)
       \delta_{ij} d x^i d x^j.
   \label{metric-0PN}
\eea Although $\alpha$ (thus $U$) contains the perturbed Newtonian
gravitational potential, the full Einstein's equation demands to
keep $\varphi$ (thus $V$) which can be regarded as the PN order
\cite{Hwang-Noh-2013b}, see equation (\ref{metric-PN}). Thus, as the
0PN limit we identify \bea
   & & \alpha = - {1 \over c^2} U, \quad
       \varphi = {1 \over c^2} V, \quad
       \chi_i = 0, \quad
       \overline{v}_i = {\bf v},
\eea and take the leading order terms in the $1/c$ expansion.
Equation (\ref{eq5}) gives $\varphi = - \alpha$, thus $V = U$.
Equation (\ref{eq3}) gives \bea
   & & \kappa = - {12 \pi G \over c^2} a \Delta^{-1}
       \left( \widetilde \varrho \overline{v}^i \right)_{,i},
\eea equation (\ref{eq6}), equation (\ref{eq7}), equation
(\ref{eq4}), and equation (\ref{eq1}), respectively, give \bea
   & & \dot {\widetilde \varrho} + 3 {\dot a \over a} \widetilde \varrho
       + {1 \over a} \left( \widetilde \varrho \overline{v}^i \right)_{,i} = 0,
   \label{mass-conservation-N} \\
   & & {1 \over a} \left( a \overline{v}_i \right)^{\displaystyle\cdot}
       + {1 \over a} \overline{v}_{i,k} \overline{v}^k
       - {1 \over a} U_{,i}
       + {1 \over a \widetilde \varrho} \widetilde p_{,i} = 0,
   \label{momentum-conservation-N} \\
   & & {\Delta \over a^2} U
       + 4 \pi G \left( \widetilde \varrho - \varrho \right) = 0,
   \label{Poisson-N} \\
   & & \dot U + {\dot a \over a} U
       - 4 \pi G a \Delta^{-1} \left( \varrho \overline{v}^i \right)_{,i}
       = 0.
   \label{eq4-0PN}
\eea Equation (\ref{eq4}) also gives equation (\ref{Poisson-N}).
Equation (\ref{eq4-0PN}) is consistent with equation
(\ref{Mom-constr-PN}). These are the well known perturbed Newtonian
hydrodynamic equations in the cosmological background; see Sections
7-9 of \cite{Peebles-1980}.

One may wonder why we need the presence of $V$ term in equation
(\ref{metric-0PN}) which is ordinarily known as the 1PN order term;
it's presence as the 1PN correction is important to have correct
light deflection in Einstein's gravity, see equation (149) in
\cite{Hwang-etal-2008}. This issue was addressed in the nonlinear perturbation
theory context in \cite{Hwang-Noh-2013b}. Here, in our PN approach
we clarify the following situation. The hydrodynamic equations and
the Poisson's equation in equations
(\ref{mass-conservation-N})-(\ref{Poisson-N}) properly follow from
equations (\ref{eq6}), (\ref{eq7}) and (\ref{eq4}), respectively,
without involving the $\varphi$ (thus $V$) term; in this sense the
0PN metric in $\alpha$ (thus $U$) is enough to get the proper
Newtonian equation. However, additionally, in order to properly have
equation (\ref{Poisson-N}) from equation (\ref{eq2}), and equation
(\ref{eq4-0PN}) from equation (\ref{eq1}) we need the presence of
the $\varphi$ (thus $V$) term which is ordinarily regarded as the
1PN correction. Therefore, although the full Newtonian equations
follow from Einstein's equation {\it without} involving the $V$
term, the presence of $V$ term is demanded by the self-consistency
of the full Einstein's equation, see
\cite{Chandrasekhar-1965,Kofman-Pogosyan-1995,Matarrese-Terranova-1996}.

As a complementary study in the perturbation theory, in
\cite{Hwang-Noh-2013b} we have shown that in two gauge conditions
(the zero-shear gauge and the uniform-expansion gauge) the fully
nonlinear perturbation equations properly reduce to the Newtonian
hydrodynamic equations in the infinite speed-of-light limit (weak
gravity, slow-motion, negligible pressure and internal energy
compared with the mass-energy density, and subhorizon limits). As
the two gauge conditions are consistent with the PN approaches, the
result is consistent with the present study based on the PN
approximation; see below equation (\ref{temporal-gauges-PN}).

%
%
\section{Discussion}

Our main point in this work is a {\it new} derivation of the 1PN
equations from the recently formulated fully nonlinear and exact
perturbation theory. In this way we have clarified the relation
between the two approaches, nonlinear perturbation theory versus PN
approach, which may play key roles in handling relativistic aspects
of nonlinear processes in cosmology. In order to clarify the
relations between the two approaches we presented some overlapping
materials with \cite{Hwang-etal-2008} and \cite{Hwang-Noh-2013a}.
The master equations and gauge strategies in both approaches are
presented assuming an ideal fluid in the presence of the
cosmological constant.

We anticipate that the fully nonlinear perturbation theory will be
useful to derive 2PN equations as well where the gravitational wave
backreaction does not appear; in the Minkowski background and in
certain gauge conditions see \cite{Chandrasekhar-Nutku-1969} for 2PN
equations, and \cite{Chandrasekhar-Esposito-1970} for 2.5PN
equations with first appearance of the radiative-reaction term.
Besides ignoring the transverse-tracefree part of the metric (which
should be handled perturbatively), one other possible limit of our
nonlinear perturbation theory in handling the higher order PN
approach is the spatial gauge condition taken leaving the spatial
part of the metric as in equation (\ref{metric-PT}). In the
perturbation theory our spatial gauge condition is the unique choice
without remnant gauge mode to all perturbation orders
\cite{Bardeen-1988,Noh-Hwang-2004,Hwang-Noh-2013a}. Whether the same
spatial gauge condition will remain suitable for higher-order PN
approximation is the subject for future investigation. Under the
current situation where even 1PN equations are not implemented in
the cosmological numerical simulation yet, we do not have
pressing demand to go for the higher PN effects, though.

We hope that at some point in near future the equations based on
these two approaches could be implemented numerically as a step
toward the relativistic numerical simulations in cosmology. The PN
approach preserves the Newtonian nature of the theory with the
Einstein's gravity contributions appearing as the PN correction
terms: see equations (\ref{E-conserv-PN})-(\ref{Mom-constr-PN}) together with the gauge condition in (\ref{Gauge-PN}), or equations (\ref{Mom-conserv-PN-2}) and (\ref{Raychaudhury-eq-2}) in a gauge-invariant form.
Thus, the 1PN equations are easy to implement in the conventional
Newtonian hydrodynamic simulation code except that it could be quite
time consuming due to the presence of bare potential terms. The 1PN
equations will be important where ${GM \over Rc^2} \sim {v^2 \over
c^2}$ is small but not negligible, and the galaxy clusters and the
large-scale cosmic structure satisfy these conditions. Although the
correction terms are small, whether the PN corrections could leave
long term (secular) signature in the large-scale structure is a
subject left for future cosmological PN numerical simulation.

The full-blown numerical relativity is often based on the ADM
(Arnowitt-Deser-Misner) equations \cite{Arnowitt-etal-1962}. As the
names of equations (\ref{eq1})-(\ref{eq7}) indicate, our nonlinear
perturbation equations are the same as the ADM equations based on
our spatial gauge condition and the perturbation theory notation; we
used the covariant conservation equations instead of the ADM
conservation equations, but the choice is a matter of convenience
and both are presented in \cite{Hwang-Noh-2013a}; for the ADM conservation equations, see equations (33) and (34) in \cite{Hwang-Noh-2013a}.
Thus, our exact and fully nonlinear perturbation equations can be regarded as exact Einstein's equations {\it adapted} to cosmological perturbation
theory. Ignoring the transverse-tracefree part of the metric in our
formulation is certainly limiting the potential applications, but as
demonstrated in \cite{Hwang-Noh-2013a}, the fully nonlinear
equations are quite powerful in deriving the higher order nonlinear
perturbation equations for the scalar- and vector-type perturbations
in diverse fundamental temporal gauge (slicing) conditions.

%
%
\section*{Acknowledgments}

We wish to thank the referee for useful suggestions.
J.H.\ was supported by KRF Grant funded by the Korean Government (KRF-2008-341-C00022).
H.N.\ was supported by grant No.\ 2012 R1A1A2038497 from NRF.

%
%

\end{document}